\documentclass[aip,apl,amsfonts,amssymb,amsmath,groupedaddress,singlecolumn]{revtex4}

\usepackage{graphicx}
\usepackage{amsmath}
\usepackage{amssymb}
\usepackage{amsfonts}
\usepackage{epstopdf}

\begin{document}

\title{Unusual thermoelectric transport anisotropy in quasi-2D, rhombohedral GeTe}

\author{Vahid Askarpour}
\affiliation{Department of Physics and Atmospheric Science, Dalhousie University, Halifax, Nova Scotia, Canada, B3H 4R2}
\author{Jesse Maassen}
\email{jmaassen@dal.ca}
\affiliation{Department of Physics and Atmospheric Science, Dalhousie University, Halifax, Nova Scotia, Canada, B3H 4R2}

\begin{abstract}
In this study, we calculate the $T$\,=\,300~K scattering and thermoelectric transport properties of rhombohedral GeTe using first-principles modeling. The room-temperature phase of GeTe has a layered structure, with cross-plane and in-plane directions oriented parallel and perpendicular to [111], respectively. Based on rigorous electron-phonon scattering, our transport calculations reveal unusual anisotropic properties; {\it n}-type GeTe has a cross-plane electrical conductivity that is roughly 3$\times$ larger than in-plane. {\it p}-type GeTe, however, displays opposite anisotropy with in-plane conducting roughly 2$\times$ more than cross-plane, as is expected in quasi-2D materials. The power factor shows the same anisotropy as the electrical conductivity, since the Seebeck coefficient is relatively isotropic. Interestingly, cross-plane {\it n}-GeTe shows the largest mobility and power factor approaching 500~cm$^2$/V-s and 32~$\mu$W/cm-K$^2$, respectively.  The thermoelectric figure-of-merit, $zT$, is enhanced as a result of this unusual anisotropy in {\it n}-GeTe since the lattice thermal conductivity is minimized along cross-plane. This decouples the preferred transport directions of electrons and phonons, leading to a threefold increase in $zT$ along cross-plane compared to in-plane. The {\it n}-type anisotropy results from high-velocity electron states formed by Ge p-orbitals that span across the interstitial region. This surprising behavior, that would allow the preferential conduction direction to be controlled by doping, could be observed in other quasi-2D materials and exploited to achieve higher-performance thermoelectrics.
\end{abstract}

\maketitle

\section{Introduction} \label{sec:intro}
Thermoelectric (TE) materials that efficiently convert thermal energy into electrical energy, or vice versa, exhibit a high TE figure-of-merit $zT=S^2 \sigma T/(\kappa^e+\kappa^L)$, where $\sigma$ is the electrical conductivity, $S$ the Seebeck coefficient, $T$ the absolute temperature, and $\kappa^{e/L}$ the electronic/lattice components of the thermal conductivity \cite{He2017}. It is well known that to increase $zT$ one must find or design materials that possess excellent electronic transport and poor lattice transport properties. One interesting class of materials that show promise are quasi-2D, or layered, crystals comprised of atomic layers that have strong interactions within each layer (intra-layer) and weak coupling between layers (inter-layer). There are both well-established (Bi$_2$Te$_3$ \cite{Goldsmid1958}) and recently discovered (SnSe \cite{Zhao2016,Chang2018}, Mg$_3$Sb$_2$ \cite{Zhang2017}, GeTe \cite{Li2018}) high-performance TEs that possess layered structures. While quasi-2D materials are actively being researched for a variety of interesting physical phenomena, a consequence of their structure is that they display anisotropic electrical and thermal properties.

The usual behavior is that conduction is largest along the in-plane direction and lowest along cross-plane, for both electrons and phonons. Recently, there have been examples where the anisotropy of layered materials can be utilized to enhance TE properties. SnSe has demonstrated $zT$\,=\,2.8 along its cross-plane direction, resulting from an ultralow $\kappa^L$ accompanied with a power factor, $PF$\,=\,$S^2\sigma$, similar to or better than that along in-plane \cite{Chang2018}. Black phosphorus has anisotropy along its two in-plane directions that is opposite for electrons and phonons (which conduct best along armchair and zigzag, respectively) \cite{Fei2014,Liu2014,Luo2015}. This anisotropic decoupling of electrons and phonons can enhance $zT$ by providing a preferential direction along which the power factor and thermal conductivity are maximized and minimized, respectively. Such ideas have motivated searches for new quasi-2D crystals with exceptional properties \cite{Zhang2016,Gorai2016}. Moreover, some layered materials have demonstrated unusual electronic structures predicted to outperform standard parabolic bands \cite{Zahid2010,Maassen2013,Wickramaratne2015}.

Here we theoretically investigate the TE properties of rhombohedral GeTe, the room temperature phase, which has a layered structure. Above $\sim$\,640~K GeTe undergoes a phase transition from a rhombohedral to a cubic (rock salt) structure. GeTe has been investigated within the context of phase-change applications \cite{Bruns2009}, ferroelectricity \cite{Polking2012} and thermoelectrics. Several theoretical \cite{Xu2011,Singh2013,Chen2013,Ding2015,Yang2016,Kagdada2018,Xing2018} and experimental \cite{Levin2013,Perumal2016,Li2017,Li2018} studies have focused on the TE characteristics of GeTe. Noteworthy examples that motivated this work include a first-principles analysis showing that the conduction states of cubic GeTe possess highly non-parabolic bands that should be beneficial for TE performance \cite{Chen2013}, and a recent demonstration of $zT$\,$\approx$\,2.4 with rhombohedral GeTe \cite{Li2018}.

In this study we utilize first-principles modeling to calculate the electron-phonon scattering rates and TE properties of rhombohedral GeTe. Our findings show that this layered material demonstrates unusual TE anisotropy: $\sigma$ for {\it n}-type GeTe is largest along the cross-plane direction, perpendicular to the atomic layers, while for {\it p}-type GeTe the trend reverses with the largest $\sigma$ oriented along in-plane. Given that $\kappa^L$ is lowest and $PF$ is largest perpendicular to the atomic layers, this decoupling of electrons and phonons via anisotropy helps boost TE performance. The source of the anisotropy is shown to arise from high cross-plane velocity conduction states. The outline of the paper is as follows: Sec. \ref{sec:theory} describes our theoretical approach and computational details, Sec. \ref{sec:results} presents our results of electron/phonon dispersions, scattering rates and TE properties, Sec. \ref{sec:discuss}-\ref{sec:conclusions} discusses and summarizes our findings.

\section{Theoretical approach} \label{sec:theory}
The electronic conductivity, $\sigma$, and Seebeck coefficient, $S$, are calculated using \cite{Scheidemantel2003,Liao2015}:
\begin{align}
\sigma_{\alpha} &= \frac{e^2}{\Omega} \sum_{\bf k} v_{\bf k}^{\alpha} v_{\bf k}^{\alpha} \tau^m_{\bf k} \left(-\frac{\partial f_{\bf k}}{\partial \epsilon_{\bf k}}\right), \label{eq:sigma} \\
S_{\alpha} &= \frac{e k_B}{\sigma_{\alpha}\Omega} \sum_{\bf k} \left[ \frac{\epsilon_{\bf k}-\epsilon_F}{k_B T} \right] v_{\bf k}^{\alpha} v_{\bf k}^{\alpha} \tau^m_{\bf k} \left(-\frac{\partial f_{\bf k}}{\partial \epsilon_{\bf k}}\right), \label{eq:seebeck}
\end{align}
where ${\bf k}$ is an electron state in the Brillouin zone (BZ), $v_{\bf k}^{\alpha}$ the electron velocity along the direction $\alpha$ ($=x,y,z$), $\tau^m_{\bf k}$ the momentum scattering time, $\epsilon_{\bf k}$ the electron energy, $f_{\bf k}$ the Fermi-Dirac distribution, $\epsilon_F$ the Fermi level and $\Omega$ the sample volume. The direction-dependent power factor is obtained using $PF_{\alpha}=S_{\alpha}^2 \sigma_{\alpha}$.

In this study we consider electron-phonon (el-ph) scattering as the dominant, intrinsic collision mechanism. The el-ph momentum scattering rates, $1/\tau^m_{\bf k}$, are computed as \cite{Liao2015}:
\begin{align}
&\frac{1}{\tau^m_{\bf k}}=\frac{2\pi}{\hbar} \sum_{\bf k'} |g({\bf k},{\bf q})|^2 \Big[(f_{\bf k'}+n_{\bf q}) \, \delta(\epsilon_{\bf k'}-\epsilon_{\bf k}-\hbar\omega_{\bf q}) \nonumber \\ 
&+ (1-f_{\bf k'}+n_{\bf q}) \, \delta(\epsilon_{\bf k'}-\epsilon_{\bf k}+\hbar\omega_{\bf q}) 
 \Big]  \Big(1-\frac{{\bf v}_{\bf k'}\cdot{\bf v}_{\bf k}} {|{\bf v}_{\bf k'}|{|\bf v}_{\bf k}|} \Big), \label{eq:itau}
 \end{align}
where ${\bf q}$ is a phonon state in the BZ, $g({\bf k},{\bf q})$\,$\propto$\,$\langle\mathbf{k}\pm\mathbf{q}|H_{\rm el-ph}|\mathbf{k}\rangle$ the el-ph coupling matrix, $\hbar \omega_{\bf q}$ the phonon energy, and $n_{\bf q}$ the Bose-Einstein distribution. The first and second term in square brackets correspond to phonon absorption and emission, respectively. In addition to conserving energy, implicit in Eq. (\ref{eq:itau}) is that el-ph processes conserve crystal momentum ${\bf k'}={\bf k}\pm{\bf q}$. The last term in parentheses has the form 1-cos($\theta$), with $\theta$ being the angle between initial $\bf k$ and final $\bf k'$ states, which captures the momentum angle change upon scattering. Small angle deflections count less than backscattering events since the forward momentum, and hence the current contribution, is not strongly altered. For clarity, the electron band ($n$), spin ($s$) and phonon branch ($\nu$) have been omitted, $|{\bf k}\rangle \rightarrow |{\bf k},n,s\rangle$ and $|{\bf q}\rangle \rightarrow |\nu,{\bf q}\rangle$, but are assumed and implicit in all summations over $\bf k$. The coupling matrix is computed including the standard and polar (Fr\"{o}hlich) el-ph interactions, with details provided in Refs. \cite{EPW1,Verdi2015,Sjakste2015}. Mobile carrier screening, at the level of Thomas-Fermi theory, is included in the calculation of $g({\bf k},{\bf q})$ and the polar component of the dynamical matrix (using same approach as in Ref. \cite{Song2017}). Screening reduces the phonon-induced potential variation, resulting in the el-ph coupling matrix being scaled by the following factor: $g({\bf k},{\bf q}) \rightarrow g({\bf k},{\bf q}) \times (qL_D)^2/[1+(qL_D)^2]$, where $L_D$ is the Debye screening length given by $1/L_D^2 = (e^2/(\varepsilon_0 \varepsilon_{\infty} k_B T)) \int D(\epsilon) f(\epsilon,\epsilon_F) [1-f(\epsilon,\epsilon_F)] \,d\epsilon$, $\varepsilon_{0/\infty}$ is the vacuum/relative high-frequency dielectric constant, and $D(\epsilon)$ is the electron density-of-states.

\subsection{Computational details}
The electron, phonon and scattering properties, required to evaluate the TE parameters, are first computed using the density functional theory (DFT) simulation package Quantum Espresso (QE) \cite{QE1,QE2}. The self-consistent electronic calculation was performed with norm-conserving relativistic pseudopotentials, GGA-PBE for exchange-correlation potential, a plane wave cutoff of 100~Ry, a uniform $\bf k$-grid of 12$\times$12$\times$12, and spin-orbit coupling. Perturbation theory, as implemented in QE, was used to compute the force constants and scattering potential due to phonons on a uniform $\bf q$-grid of 6$\times$6$\times$6, along with the Born effective charges and dielectric constants (40.88 for in-plane, 36.91 for cross-plane). The el-ph scattering rates and TE parameter calculations were performed with the EPW code \cite{EPW2}. The DFT-computed electron Hamiltonian and el-ph coupling matrix were transformed to a Wannier representation using 16 maximally localized Wannier functions as a basis \cite{Wannier}. From the Wannier representation, the electron/phonon energies and el-ph coupling matrix were interpolated back onto much finer $\bf k$- and $\bf q$-grids of 150$\times$150$\times$150 and 80$\times$80$\times$80, respectively, which served to compute the el-ph scattering rates and TE parameters. The delta functions in Eq. (\ref{eq:itau}) were approximated as gaussians with a broadening parameter of 5 meV. By performing DFT simulations of pristine/undoped GeTe, our methodology incorporates the effect of doping by shifting the Fermi level to obtain the desired carrier concentration (the rigid band approximation), and includes the effect of mobile carrier screening by scaling the el-ph potential variation (via Thomas-Fermi theory, described above). This DFT approach has been used to study a variety of materials \cite{Liao2015,Hung2017,Ma2018}, and has shown good agreement with experiment for Si \cite{Qiu2015}, GaAs \cite{Zhou2016,Liu2017}, PbTe \cite{Song2017}, among others. While ionized impurity scattering is known to play a role in doped semiconductors \cite{Qiu2015}, this study focuses on the transport properties of pristine, defect-free GeTe with el-ph collisions as the intrinsic scattering mechanism.

\begin{figure}	
\includegraphics[width=3.3in]{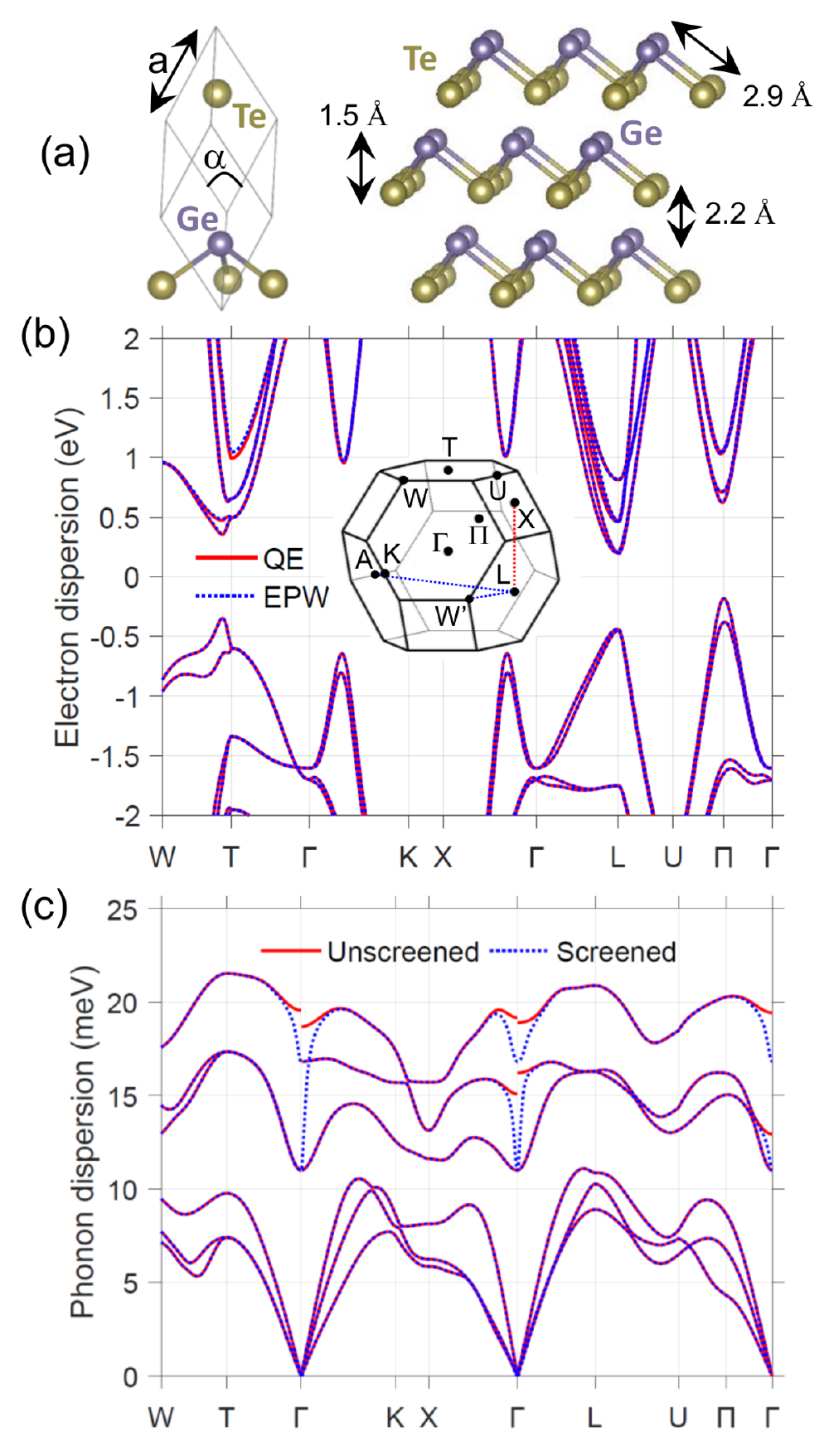}
\caption{(a) Primitive cell and atomic structure of rhombohedral GeTe. Electron (b) and phonon (c) dispersions along high-symmetry lines in the BZ. Point $\Pi$ is located at (0.2118, 0.3714, 0.1918) in reduced coordinates, and is the location of the VBM. Unscreened (red solid) and screened (blue dashed) phonon energies are compared for $n$\,=\,5$\times$10$^{19}~$cm$^{-3}$ and $T$\,=\,300~K.} \label{fig:bands}
\end{figure}

\section{Results} \label{sec:results}
\subsection{Atomic structure} 
At $T$=300~K, GeTe crystallizes into a rhombohedral ({\it R}3{\it m}) structure with two atoms in the primitive cell, as shown in Fig.~\ref{fig:bands}(a). In this phase GeTe is a layered, quasi-2D material comprised of two-atom-thick sheets with strong intra-layer and weak inter-layer coupling, and with a cross-plane direction pointing along [111]. Each Ge and Te atom has three nearest neighbors (within its layer), with a Ge-Te bond length of 2.87~\AA\, (3.26~\AA\, between adjacent layers). The thickness of each two-atom-thick layer is 1.49~\AA\,, and the inter-layer distance is 2.16~\AA. The DFT-optimized lattice parameters are $a$\,=\,4.395~\AA\,, $\alpha$\,=\,57.72$^\circ$, which are in reasonable agreement with the experimental values of $a$\,=\,4.288~\AA\, and $\alpha$\,=\,57.93$^\circ$~\cite{Pereira2013} and to other DFT calculations \cite{Campi2017,Kagdada2018}. Ref. \cite{Campi2017} studied the role of van der Waals (vdW) interaction and found that GGA-PBE (without vdW) provided atomic coordinates in better agreement with experiment, and that for a given atomic configuration the phonon properties were relatively insensitive to the choice in DFT functional. The location (in terms of the lattice vectors) of the Ge and Te atoms are ($\gamma$,$\gamma$,$\gamma$) and ($-\gamma$,$-\gamma$,$-\gamma$), respectively, where $\gamma$\,=\,0.2348. Above $\sim$\,640~K \cite{Chattopadhyay1987,Levin2013}, GeTe takes a high-symmetry cubic phase (rock-salt structure), where the angle between the primitive vectors is 60$^\circ$ and each atom has six equivalent nearest neighbors. Below $\sim$\,640~K, there is a continuous transition from cubic to rhombohedral, where the angle between vectors drops below 60$^\circ$ and one atomic species moves slightly along the [111], resulting in each atom having three nearest neighbors.

\subsection{Electron and phonon dispersions}
The starting point for carrying out scattering and transport calculations are accurate descriptions of the electron and phonon dispersions. Fig.~\ref{fig:bands}(b) shows the electron band structure along high-symmetry lines in the BZ (zero energy corresponds to mid-gap). Our electron dispersion is similar to previous calculations \cite{DiSante2013,Li2018,Xing2018,Kagdada2018}. GeTe has an indirect band gap of $E_g$\,=\,0.36~eV, consistent with previous theoretical studies \cite{Singh2013,DiSante2013}, which is less than the measured $E_g$\,=\,0.55~eV \cite{Shportko2008}. The smaller theoretical $E_g$ does not impact our transport calculations, since the Fermi level is always close enough to either the valence or conduction band to avoid bipolar conduction. The conduction band minimum (CBM) is located slightly off the L point and has a valley degeneracy of three, $g_v$\,=\,3. A closer view of the conduction edge shows that the states are split as a result of spin-orbit interaction, resulting in non-parabolic Rashba states \cite{DiSante2013}. There are higher-energy secondary bands near T ($g_v$\,=\,1) and L ($g_v$\,=\,3). The valence band maximum (VBM) occurs slightly off the $\Gamma$-U line at $\Pi$, which is six-fold degenerate ($g_v$\,=\,6). There are secondary valence bands located near T and L. A comparison of the band structure calculated with QE (red solid line) and EPW (blue dashed line) shows good agreement, indicating the Wannier orbital basis accurately represents the electronic system.

Fig.~\ref{fig:bands}(c) presents the phonon dispersion, containing three acoustic and three optical branches (2 atoms in the primitive cell). Our phonon dispersion is similar to previous calculations \cite{Campi2017,Kagdada2018}. The maximum phonon energy of 22 meV is relatively small, owing to the large mass and size of Te, and is comparable to other soft-bond and heavy materials such as Au or Bi$_2$Te$_3$. The long-range polar el-ph interaction leads to a splitting of the optical modes at the zone center (unscreened case), resulting from significant charge transfer as the atoms oscillate. This effect is most pronounced near $\bf q$\,$\approx$\,0, where the phonon wavelength is long. Mobile carrier screening reduces the polar interaction; with $n$\,=\,5$\times$10$^{19}$~cm$^{-3}$ the polar interaction is significantly weakened (screened case), as also seen in the case of PbTe \cite{Song2017}. From the slope of the longitudinal acoustic bands we extract the sound velocity: along $\Gamma$-K (in-plane) and $\Gamma$-T (cross-plane), we find 3260 m/s and 2670 m/s, respectively. The in-plane group velocity is larger than that of cross-plane, as expected in quasi-2D materials. We note, however, that this anisotropy is not as pronounced as with other layered materials, such as graphite or MoS$_2$, where the phonon dispersion is rather flat along the cross-plane direction. This suggests, along with a relatively small inter-layer distance, that the inter-layer coupling is not as weak as with other quasi-2D materials.

\begin{figure}	
\includegraphics[width=3.3in]{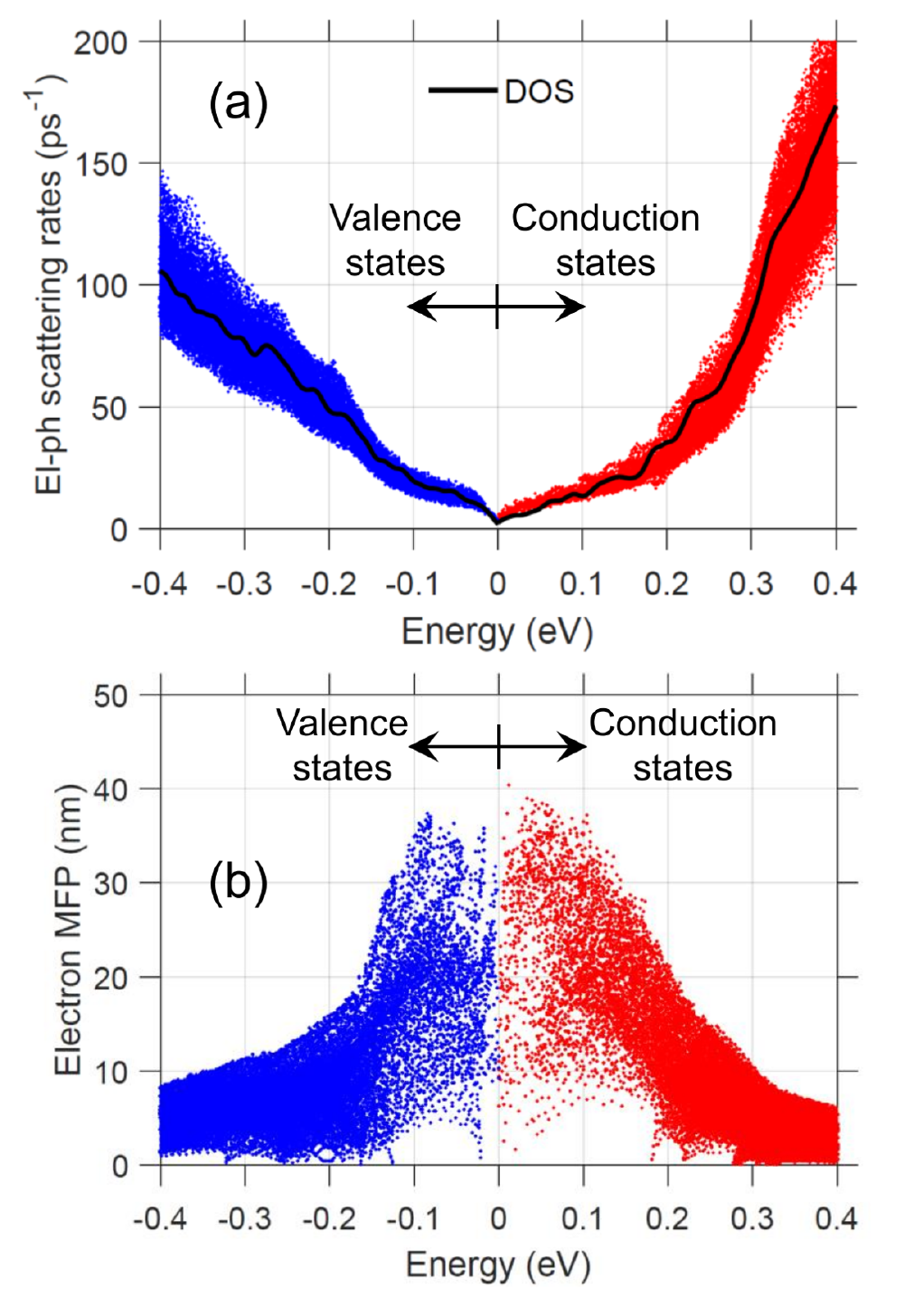}
\caption{(a) Screened momentum scattering rates, $1/\tau^m_{\bf k}$, and (b) mean-free-paths (MFP), $l_{\bf k}=|v_{\bf k}|\tau^m_{\bf k}$, versus energy relative to the band edges for {\it n}-GeTe (1.5$\times$10$^{19}$~cm$^{-3}$) and {\it p}-GeTe (3.0$\times$10$^{19}$~cm$^{-3}$). The selected electron and hole concentrations maximize the power factor. $T$\,=\,300~K.} \label{fig:rate_mfp}
\end{figure}

\subsection{Electron-phonon scattering rates and mean-free-paths}
Next, using our computed electron and phonon dispersions, along with the el-ph coupling matrix, we can evaluate the scattering rates. The screened momentum scattering rates for $T$\,=\,300~K are shown in Fig.~\ref{fig:rate_mfp}(a). Each dot corresponds to the scattering rate, $1/\tau^m_{\bf k}$, of an electron/hole state in the BZ. The scattering rates increase with carrier energy, as they get further away from the band edges. Physically this makes sense; at higher energies there are more available final states for electrons to scatter into. This is confirmed by comparing $1/\tau^m_{\bf k}$ to the electron density-of-states (DOS), which show good agreement. A scattering rate that follows the DOS is expected in the case of a parabolic band with a deformation potential treatment of el-ph scattering \cite{LundstromBook}, but here is found to be valid for a more complex band structure and rigorous treatment of scattering. We note, however, that polar optical phonon scattering should result in $1/\tau^m_{\bf k}$ that does not necessarily follow the DOS, and that is roughly constant or decreasing with increasing carrier energy \cite{LundstromBook,Cao2018}. Unscreened el-ph scattering calculations confirm that polar phonon scattering obeys the expected trend and is dominant (see Appendix \ref{app:unscreened}). Screening is found to significantly reduce the scattering rates near the band edges resulting in $1/\tau^m_{\bf k}$\,$\propto$\,DOS, with roughly similar contributions from all phonon branches. There are visible upticks in $1/\tau^m_{\bf k}$ near CBM+0.2~eV and VMB-0.15~eV, due to higher-energy secondary bands.

The mean-free-path (MFP), calculated using $l_{\bf k}=|v_{\bf k}| \tau^m_{\bf k}$, is shown in Fig.~\ref{fig:rate_mfp}(b). Note that we focus on the states within 0.4~eV of the band edges, since carrier transport outside of this range is negligible. The MFPs are found to be energy-dependent, reaching $\sim$35~nm near the band edges and decaying to $\sim$5~nm at higher energy. This trend arises because $1/\tau^m_{\bf k}$ increases more rapidly with energy than the electron velocities. (A single parabolic band with deformation potential el-ph scattering predicts a constant MFP \cite{LundstromBook}.)

In the literature it is common to find theoretical studies of TE transport that rely on DFT for accurate descriptions of electron and phonon dispersions, but adopt simpler models for the scattering physics, such as a constant scattering time ($\tau^m_{\bf k}$\,=\,$\tau_0$) or a constant MFP ($l_{\bf k}$\,=\,$l_0$). Our rigorous el-ph scattering calculations show that, in the case of GeTe, both the scattering rates and MFP are energy-dependent. An alternative simple model, that appears to work well, assumes the scattering rate is proportional to the electron DOS \cite{Witkoske2017,Wang2018}.

\subsection{Thermoelectric transport properties}
\begin{figure}	
\includegraphics[width=6in]{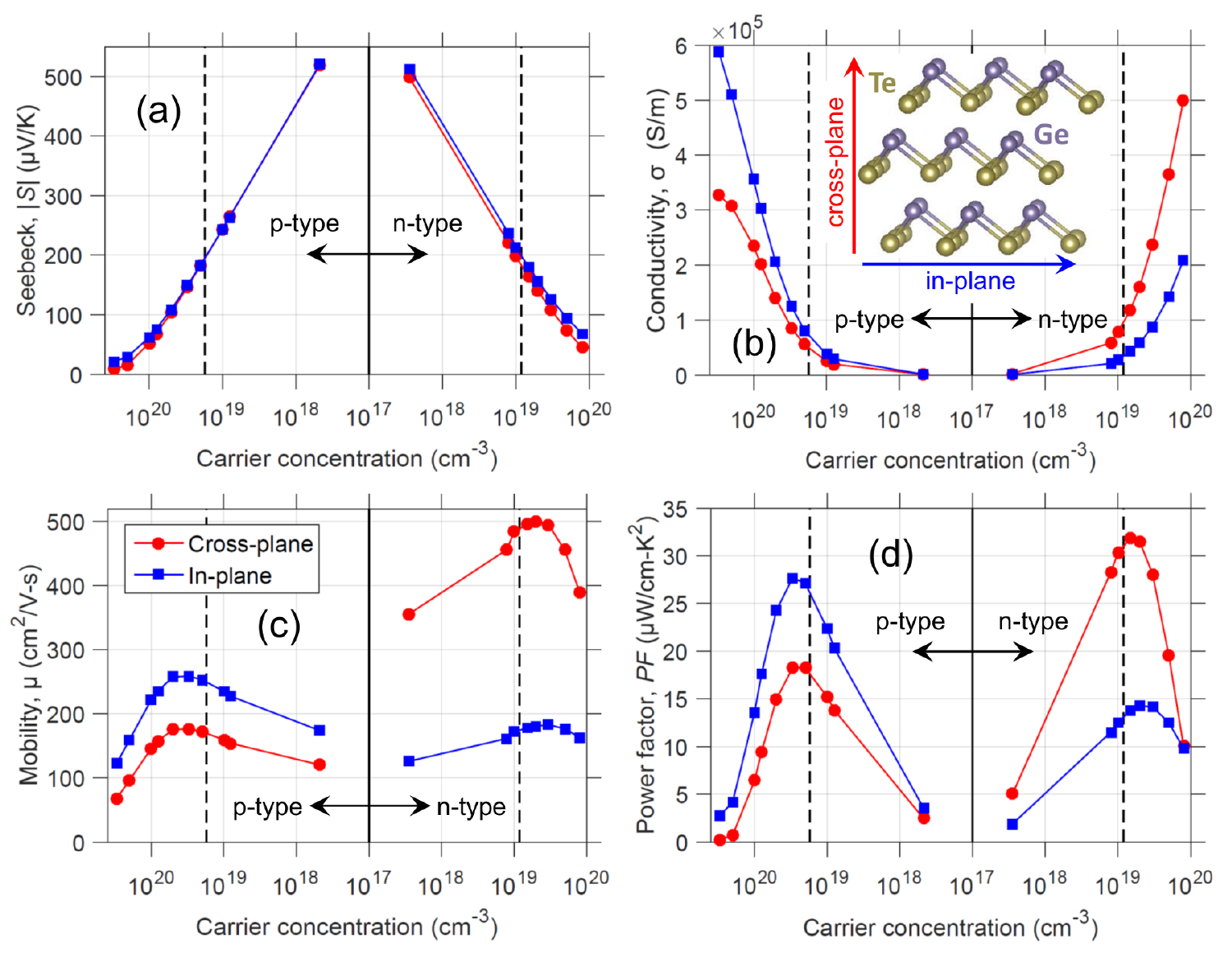}
\caption{Carrier concentration dependence of Seebeck coefficient (a), electrical conductivity (b), mobility (c) and power factor (d). The cross-plane and in-plane TE results are plotted as red circles and blue squares, respectively. Vertical black dashed lines denote the VBM and CBM. $T$\,=\,300~K.} \label{fig:TE_param}
\end{figure}

With the calculated dispersions and scattering properties, we can assess the TE transport characteristics of rhombohedral GeTe. The Seebeck coefficient, $S$, shown in Fig.~\ref{fig:TE_param}(a), demonstrates standard behavior with $|S|$ decreasing with increasing carrier concentration. The trend of $|S|$ versus log($n,p$) is linear when the Fermi level $\epsilon_F$ is inside the band gap (and several $k_BT$ away from the band edge), and sub-linear when $\epsilon_F$ moves inside the band. $|S|$\,$\approx$\,200~$\mu$V/K at the band edges (black dashed lines), which is similar to other good thermoelectric materials, for example bulk Bi$_2$Te$_3$ \cite{Maassen2013}. For a fixed carrier concentration, {\it p}-type $|S|$ is slightly larger than {\it n}-type, owing to the larger DOS of the valence states. The Seebeck is nearly isotropic, when comparing the values along the cross-plane and in-plane directions. Note that the two in-plane directions (binary and bisectrix axes) give the same TE values, thus we only show the in-plane and cross-plane (trigonal axis) directions.

Contrary to $S$ the electrical conductivity, $\sigma$, shows pronounced anisotropy in Fig.~\ref{fig:TE_param}(b). {\it p}-GeTe shows a larger in-plane conductivity ($\sigma_{||}$) compared to cross-plane conductivity ($\sigma_{\perp}$). $\sigma_{||}$\,$>$\,$\sigma_{\perp}$ is the expected, and commonly observed, anisotropic behavior in quasi-2D materials because the atomic interactions are strongest in-plane and weakest cross-plane. Surprisingly, {\it n}-type GeTe reverses the typical anisotropy of layered materials with larger cross-plane conduction, $\sigma_{\perp}$\,$>$\,$\sigma_{||}$. The anisotropy ratio reaches $\sigma_{\perp}/\sigma_{||}$\,$\approx$\,2.8 with {\it n}-GeTe, compared to $\sigma_{||}/\sigma_{\perp}$\,$\approx$\,1.8 with {\it p}-GeTe. The conductivity values of GeTe are high. As a comparison, SnSe has a conductivity reaching $\sim$1.6$\times$10$^5$~S/m for $p$\,=\,4$\times$10$^{19}$~cm$^{-3}$ at room temperature \cite{Zhao2016} (which led to ultrahigh power factor values). For the same carrier concentration, GeTe has $\sigma_{p,||}$\,=\,1.7$\times$10$^5$~S/m and $\sigma_{n,\perp}$\,=\,3.0$\times$10$^5$~S/m. This tells us the carrier mobility of GeTe is similar to or larger than that of SnSe. Fig.~\ref{fig:TE_param}(c) presents the carrier mobility, $\mu_n=\sigma_n/(en)$ and $\mu_p=\sigma_p/(ep)$. $\mu$ shows the same carrier-type-dependent anisotropy as $\sigma$, as expected, with $\mu_{n,\perp}$ reaching 500 cm$^2$/V-s and $\mu_{p,||}$ approaching only 260 cm$^2$/V-s. This suggests that the velocity of the conduction states are larger along the cross-plane direction. The mobility peaks near the band edges since at higher carrier concentrations the velocity of the states decreases and at lower carrier concentrations incomplete screening results in more scattering. 

The power factor, $PF$\,=\,$S^2\sigma$, shown in Fig.~\ref{fig:TE_param}(d), demonstrates the same unusual anisotropy as $\sigma$. Interestingly, the maximum $PF$ of 32~$\mu$W/cm-K$^2$ is obtained with {\it n}-GeTe along the cross-plane at $n$\,=\,1.5$\times$10$^{19}$~cm$^{-3}$ ($PF_{\perp}$/$PF_{||}$\,=\,2.3 at maximum $PF$). With {\it p}-GeTe, the largest $PF$ is 28~$\mu$W/cm-K$^2$ along in-plane at $p$\,=\,3.0$\times$10$^{19}$~cm$^{-3}$ ($PF_{||}$/$PF_{\perp}$\,=\,1.5 at maximum $PF$). The $PF$ peak occurs when the Fermi level is slightly inside the conduction and valence bands. The largest $PF$ anisotropy ratio for each carrier type is $PF_{\perp}$/$PF_{||}$\,=\,2.7 at $n$\,=\,3.5$\times$10$^{17}$~cm$^{-3}$, and $PF_{||}$/$PF_{\perp}$\,=\,11.4 at $p$\,=\,3$\times$10$^{20}$~cm$^{-3}$. 


Quasi-2D materials very often show anisotropy in the form of lower cross-plane conduction compared to in-plane (both electrical and thermal), including Bi$_2$Te$_3$, MoS$_2$, graphite, SnSe and black phosphorus, among others. It is unusual, however, to find a layered material with anisotropy that {\it i)} prefers cross-plane conduction, {\it ii)} can be reversed by changing the carrier type and {\it iii)} is opposite for electrons and phonons.

\begin{figure}	
\includegraphics[width=6in]{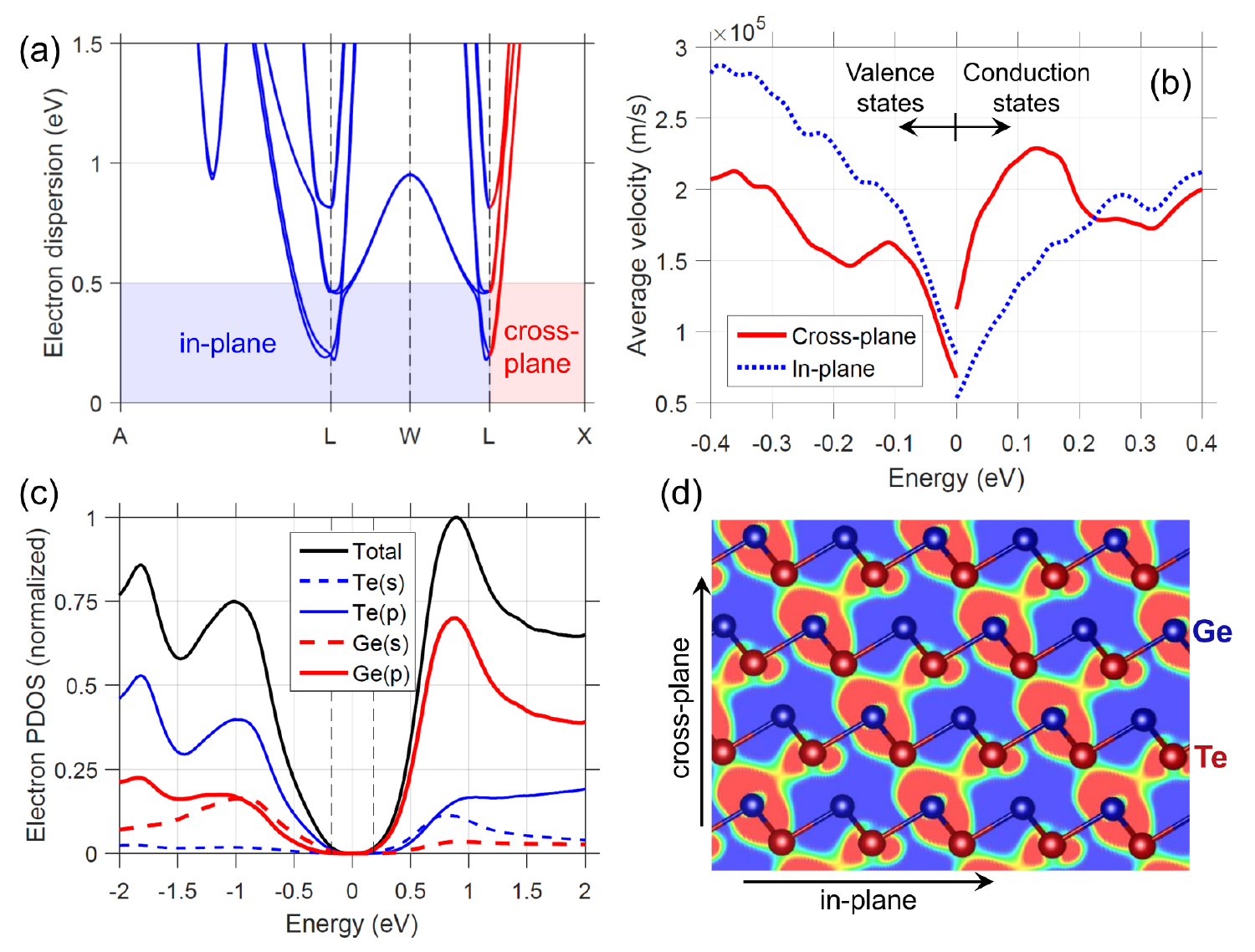}
\caption{(a) Electron dispersion of conduction states, along in-plane (blue) and cross-plane (red) directions, as indicated by the dashed lines in the inset of Fig.~\ref{fig:bands}(b). (b) Average band velocity along cross-plane (red solid) and in-plane (blue dashed) directions. See text for definition of average velocity. (c) Orbital-resolved DOS versus energy, with zero energy corresponding to mid-gap. (d) Atomic structure and charge density of the conduction states within 50~meV of the CBM.} \label{fig:charge_velo}
\end{figure}

To understand why {\it n}-GeTe has large cross-plane transport, Fig.~\ref{fig:charge_velo}(a) shows the electron dispersion along several in-plane directions and the cross-plane direction (indicated as dashed lines in the inset of Fig.~\ref{fig:bands}(b)). Focusing on the conduction band edge, near L, the in-plane directions show splitting of the bands due to spin-orbit coupling (Rashba effect), while along the cross-plane direction the bands do not split and have strong curvature. The velocity of each state is related to the curvature of the bands, $\hbar v_{\bf k}^{\alpha}=\partial \epsilon_{\bf k}/\partial k_{\alpha}$ ($\alpha$=$x$,$y$,$z$). Fig.~\ref{fig:charge_velo}(b) presents the average velocity versus energy, $\langle v^{\alpha}(\epsilon)\rangle = \sum_{\bf k} |v_{\bf k}^{\alpha}| \delta(\epsilon-\epsilon_{\bf k})/\sum_{\bf k} \delta(\epsilon-\epsilon_{\bf k})$, which demonstrates that the velocity along cross-plane is higher than in-plane near the CBM. The in-plane velocity is larger near the VBM. (The two in-plane velocities are very similar, and are averaged in Fig.~\ref{fig:charge_velo}(b).) The next question is why do we obtain high cross-plane velocity conduction states? According to the orbital-resolved DOS, in Fig.~\ref{fig:charge_velo}(c), the conduction states are mainly comprised of Ge $p$-orbitals (mostly $p_z$) and Te $s$-orbitals. The strong overlap of the Ge $p$-orbitals results in a channel of delocalized charge that spans across the atomic layers, as seen in Fig.~\ref{fig:charge_velo}(d) presenting the charge density of the CBM within a 50~meV energy range. Significant charge density is observed in the interstitial region, representing a channel for electron flow across the layers. The charge distribution of the valence states, compared to the conduction states, shows less charge in the interstitial region.

\section{Discussion} \label{sec:discuss}
The unusual anisotropy predicted for {\it n}-GeTe, discussed above, is ideal for thermoelectric performance because the anisotropy for electron and phonon transport is opposite: $\sigma_{\perp}$\,$>$\,$\sigma_{||}$ and $\kappa_{||}^L$\,$>$\,$\kappa_{\perp}^L$. First-principles calculations give $\kappa_{||}^L$\,=\,2.9~W/m-K and $\kappa_{\perp}^L$\,=\,2.0~W/m-K \cite{Campi2017}. Thus, the TE figure-of-merit, $zT=S^2 \sigma T / (\kappa^e + \kappa^L)$, is maximized along the cross-plane direction since $\sigma_{\perp}$ and $\kappa_{\perp}^L$ reach their maximum and minimum values, respectively. Using the predicted $\kappa_{\perp,||}^L$, the optimal room-temperature $zT$ values are $zT^n_{\perp}$\,=\,0.39 ($\kappa^{e,n}_{\perp}$\,=\,0.32 W/m-K), $zT^n_{||}$\,=\,0.13 ($\kappa^{e,n}_{||}$\,=\,0.30 W/m-K), $zT^p_{\perp}$\,=\,0.25 ($\kappa^{e,p}_{\perp}$\,=\,0.22 W/m-K) and $zT^p_{||}$\,=\,0.25 ($\kappa^{e,p}_{||}$\,=\,0.36 W/m-K) -- the carrier concentration that maximizes $zT$ is 2$\times$10$^{19}$~cm$^{-3}$, except for $zT^n_{\perp}$ at 10$^{19}$~cm$^{-3}$. The anisotropic ratios are $zT^n_{\perp}$/$zT^n_{||}$\,=\,3 and $zT^p_{||}$/$zT^p_{\perp}$\,=\,1, thus illustrating the benefit with {\it n}-GeTe. {\it p}-GeTe shows no benefit since both electron and phonon transport are at their lowest along cross-plane.

While the maximum $PF$ of rhombohedral GeTe is relatively large ($>$\,$30$~W/cm-K$^2$), $zT$ is underwhelming due to the relatively large $\kappa^L$. This can be improved, for example, by alloying to reduce $\kappa^L$ which was successfully adopted in Ref. \cite{Li2018}. This decoupling of maximum electron and phonon transport directions has been observed in another quasi-2D material black phosphorus \cite{Fei2014,Liu2014,Luo2015}, however in that case the anisotropy was along two in-plane directions (armchair and zigzag). Another interesting point is that GeTe is predicted to have better {\it n}-type thermoelectric properties. There is a need to discover more high-performance {\it n}-type TE materials, to complement the {\it p}-type materials in TE couples, and there have been several recent advances demonstrated with, for example, SnSe \cite{Chang2018} and Mg$_3$Sb$_2$ \cite{Zhang2017}. SnSe, in particular, shares similarities with GeTe -- they are both IV-VI semiconductors and have layered structures with relatively small inter-layer spacing. Interestingly, SnSe has demonstrated relatively large cross-plane conduction (as large as in-plane) when doped n-type \cite{Chang2018}. DFT studies of n-SnSe \cite{Ma2018,Kutorasinski2015} show that the lowest conduction band results in predominantly in-plane transport, but that a secondary higher-energy band ($\sim$0.1~eV) displays large cross-plane velocity that may lead to $\sigma_{\perp}>\sigma_{||}$, as we observe with the conduction band of GeTe.

Comparing our calculated results with experimental data, we observe significant differences. Samples are always {\it p}-type, typically heavily-doped ($>$10$^{20}$~cm$^{-3}$) and exhibit larger $PF$ than predicted by our model (for those $p$ values). Studies have measured room-temperature $PF$ values around 6-9~$\mu$W/cm-K$^2$ for $p$\,=\,6-9$\times$10$^{20}$~cm$^{-3}$ \cite{Levin2013}, and have reached $\approx$\,15~$\mu$W/cm-K$^2$ \cite{Li2017} and $\approx$\,25~$\mu$W/cm-K$^2$ \cite{Li2018} near $p$\,=\,1.5-2$\times$10$^{20}$~cm$^{-3}$ (for this $p$ range we find at most 4 $\mu$W/cm-K$^2$). Our results predict a maximum {\it p}-type $PF$ of 28~$\mu$W/cm-K$^2$ around $p$\,=\,3$\times$10$^{19}$~cm$^{-3}$, while experimentally the optimal $p$ is near $p$\,=\,2$\times$10$^{20}$~cm$^{-3}$ \cite{Li2017}. A closer comparison between our results and measured data reveals that, in the range of $p$\,=\,1-3$\times$10$^{20}$~cm$^{-3}$, the experimental $S$ and $\sigma$ are roughly 2-4$\times$ larger and 2-3$\times$ smaller than theory, respectively. While measured $\sigma$ is less than our calculations, the larger $S$ results in a larger experimental $PF$, since $PF$ depends on $S^2$. We also note that there is variability among the measured thermoelectric data, likely due to differences in GeTe samples.

There are several possible explanations for the discrepancy between theory and experiment: {\it i)} GeTe is always {\it p}-type due to a high density ($>$10$^{20}$~cm$^{-3}$) of native Ge vacancy defects \cite{Edwards2005,Huang2012}. At such large concentrations, the defects may alter the electronic structure of GeTe such that it becomes distinct from pristine GeTe, and/or introduce significant ionized point defect scattering. {\it ii)} It is common to alloy GeTe with other elements such as Pb and Bi \cite{Li2017,Li2018} (up to $\sim$10\%), which could also change the electronic structure. {\it iii)} The samples are often polycrystalline with many grain boundaries potentially introducing grain boundary scattering, and displaying angle-averaged properties that would prevent an observation of the predicted electron anisotropy. {\it iv)} Hall effect measurements are often utilized to extract the Hall concentration, which can be different from the actual carrier concentration and may represent a source of error \cite{LundstromBook}. Due to the aforementioned native defects in GeTe, it will be important to find strategies to block or compensate the Ge vacancies to achieve {\it n}-type GeTe and observe its predicted anisotropy.

\section{Conclusions} \label{sec:conclusions}
First-principles modeling was utilized to compute the electron-phonon (el-ph) scattering rates and thermoelectric transport properties of rhombohedral GeTe at $T$\,=\,300~K. In this phase, GeTe is a quasi-2D material with atomic layers oriented perpendicular to the [111] direction and an inter-layer distance of 2.2~\AA. The electron/phonon dispersions and el-ph scattering rates were computed and analyzed. While a constant scattering time or mean-free-path is commonly adopted for TE transport calculations, our results indicate that both these quantities are energy dependent. Assuming a scattering rate proportional to the electron DOS is found to work well as a simple scattering model. The transport characteristics display a pronounced and unusual anisotropy; the electrical conductivity, $\sigma$, is largest along the cross-plane direction with {\it n}-type GeTe ($\sigma_{\perp}$/$\sigma_{||}$\,=\,2.8), and is largest along in-plane with {\it p}-type GeTe ($\sigma_{||}$/$\sigma_{\perp}$\,=\, 1.8). Thus, the preferential conduction direction can be tuned with doping.

With a relatively isotropic Seebeck coefficient, $S$, the power factor, $PF$\,=\,$S^2\sigma$, demonstrates the same anisotropy as $\sigma$ and reaches a maximum of 32~$\mu$W/cm-K$^2$ (cross-plane) with {\it n}-GeTe and 28~$\mu$W/cm-K$^2$ (in-plane) with {\it p}-GeTe. Since the lattice thermal conductivity is lowest along the cross-plane direction, {\it n}-GeTe demonstrates opposite anisotropy for electrons and phonons that is beneficial for TE performance. We estimate that the anisotropy in the TE figure-of-merit $zT$ is $\approx$\,3 for {\it n}-type and $\approx$\,1 (no benefit) for {\it p}-type. This unusual anisotropy is explained by analyzing the electron conduction states, which show spin-split Rashba bands in-plane with high cross-plane velocity due to strong coupling of the Ge p$_z$-orbitals leading to significant charge in the interstitial region.

While the $PF$ values are relatively high, $zT$\,$<$\,1  because of the large lattice thermal conductivity, which could benefit from alloying or nanostructuring. Challenges to experimentally observing this predicted anisotropy include producing {\it n}-GeTe, as GeTe is always heavily {\it p}-type due to intrinsic defects, and making single crystal samples. This interesting behavior, however, could also be found in other chemically-similar quasi-2D materials, potentially leading to improvements in TE performance.

\appendix
\section{Role of screening on el-ph scattering}
\label{app:unscreened}
\begin{figure}	
\includegraphics[width=3in]{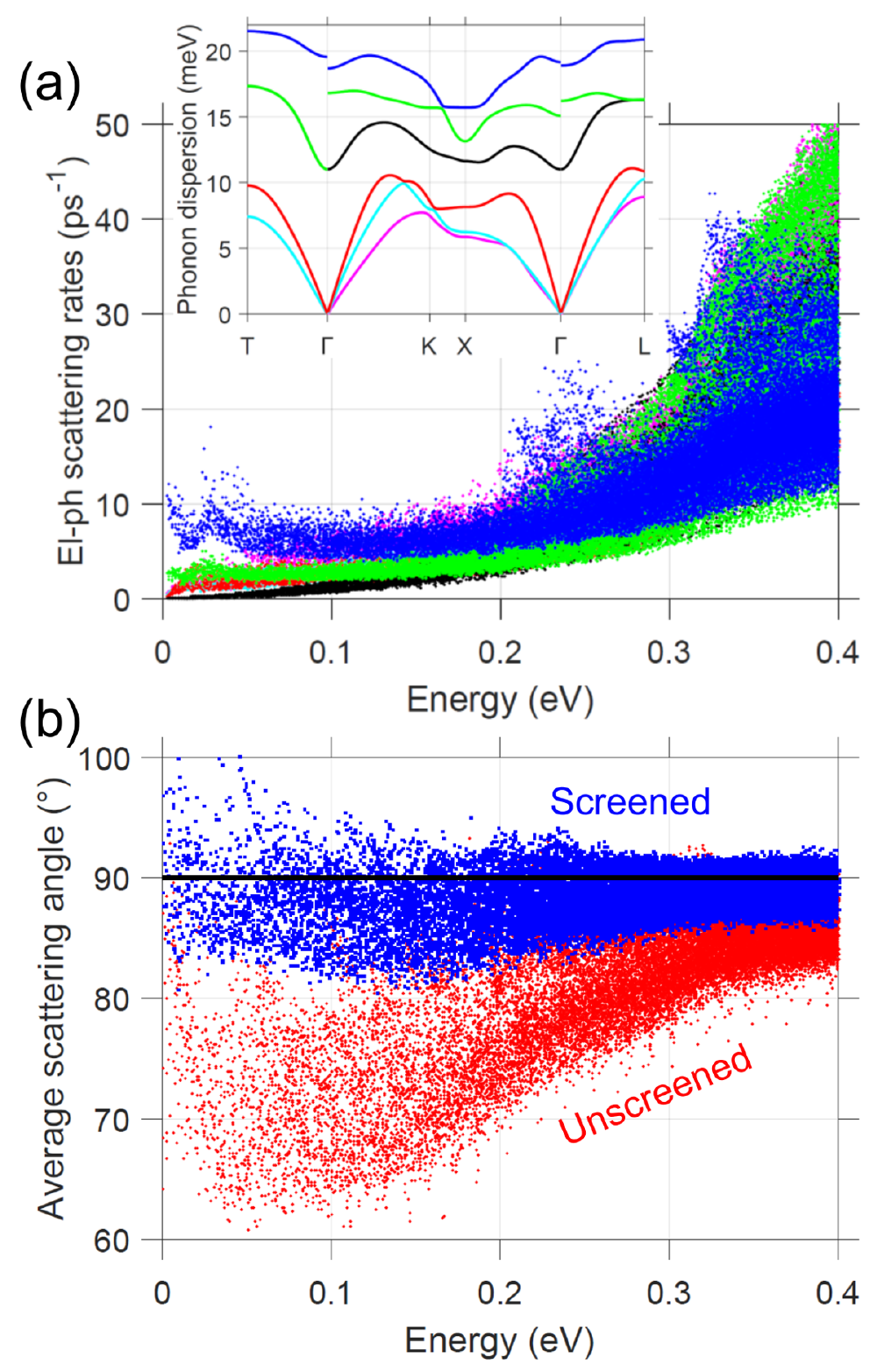}
\caption{Unscreened, phonon-resolved momentum scattering rates (a) and average scattering angle (b) of conduction states versus energy relative to the CBM, for $n$\,=\,5$\times$10$^{19}$~cm$^{-3}$ and $T$\,=\,300~K. The scattering rates are resolved according to the different phonon branches indicated in the inset showing the unscreened phonon dispersion.} \label{fig:unscreened_rate}
\end{figure}

As a comparison to the screened $1/\tau^m_{\bf k}$ shown in Fig.~\ref{fig:rate_mfp}, Fig.~\ref{fig:unscreened_rate}(a) presents the unscreened, phonon-resolved el-ph momentum scattering rates for the conduction states of GeTe. The scattering rates resulting from each phonon branch, shown in the inset, are presented separately. Without screening polar optical phonon scattering, arising from the highest-energy phonon branch, dominates for energies near the band edge. This polar phonon scattering is roughly constant in energy (or slightly decreasing) below 0.2~eV. One can observe the onset of optical phonon emission near 20~meV, corresponding to a quick rise in scattering rates. Screening reduces scattering near the band edge resulting in an energy dependence that closely matches the electron DOS, with similar contributions from all phonon branches. 

Fig.~\ref{fig:unscreened_rate}(b) shows the average scattering angle (angle between initial and final velocity after scattering event). This quantity is calculated using the definition of scattering time, $\tau_{\bf k}$, which is given by Eq.~(\ref{eq:itau}) but without the factor in parentheses that depends on velocity. The ratio of the scattering time over the momentum scattering time defines the average $\langle \cos \theta \rangle=1-\tau_{\bf k}/\tau^m_{\bf k}$, where $\theta$ is the angle between initial and final velocity. Fig.~\ref{fig:unscreened_rate}(b) presents $\cos^{-1}(1-\tau_{\bf k}/\tau^m_{\bf k})$, which we interpret as the average scattering angle. In the unscreened case, polar optical phonon scattering on average results in small angle deflections $<90^{\circ}$, as would be expected \cite{LundstromBook}. Screening, however, brings the average scattering angle closer to $90^{\circ}$, which corresponds to isotropic scattering (equal probability of scattering in any direction). Thus, screening is found to alter both the energy-dependence and angle-dependence of the el-ph scattering characteristics.

\begin{acknowledgments}
This work was partially supported by DARPA MATRIX (Award No. HR0011-15-2-0037) and NSERC (Discovery Grant RGPIN-2016-04881), with computational resources provided by Compute Canada. 
\end{acknowledgments}

\end{document}